\pdfoutput=1
\documentclass[12pt,onecolumn]{IEEEtran}
\usepackage{latexsym}
\usepackage{amsmath,amssymb,graphicx}
\usepackage{color}
\usepackage{citesort}
\newcommand{\Qv}{\vec{Q}}
\newcommand{\Pv}{{\vec{P}}}
\newcommand{\Xv}{\vec{X}}

\newcommand{\xv}{{\vec{x}}}
\newcommand{\yv}{{\vec{y}}}

\usepackage{verbatim}

\newtheorem{thm}{Theorem}[section]
\newtheorem{lemma}{Lemma}[section]

\renewcommand{\baselinestretch}{1.3}

\begin{document}
\title{On the Thermodynamic Temperature of a General Distribution}
\author{\authorblockN{Krishna R. Narayanan and Arun R. Srinivasa}
\\\authorblockA{Texas A \& M University, College Station,
TX 77843 \\ krn@ece.tamu.edu,asrinivasa@tamu.edu}
}
\maketitle
\begin{abstract}

The concept of temperature is one of the key ideas in describing the thermodynamical
properties of a physical system. In classical statistical mechanics of ideal gases, the notion
of temperature can be described in two different ways, the kinetic temperature and the
thermodynamic temperature. For the Boltzmann distribution, the two notions lead to the same
result. However, for a general probability density function, while the kinetic temperature has
been commonly used, there appears to be no corresponding general definition of thermodynamic
temperature. In this paper, we propose such a definition and show that it is connected to
the Fisher information associated with the distribution of the momenta.
\end{abstract}

\newpage

\section{Introduction}

The inverse of the thermodynamic temperature of a system in equilibrium can be defined as the rate of change of entropy
with energy \cite{callen}. It can also be thought of as a measure of the kinetic energy of the particles composing the
system. However, for a system that is not in equilibrium, while it is possible to define a temperature
using the kinetic energy of the system, there appears
to be no commonly accepted notion of thermodynamic temperature.

In this paper, we introduce a definition of thermodynamic temperature for
any  probability density function (PDF) on the momentum of the particles.
Our main contribution in this paper is to consider a particular
form of perturbation of the momentum that increases the entropy and the energy associated with the distribution.
This perturbation can be thought of a statistical realization of ``heating". Then, we
 define the thermodynamic temperature
as the rate of change of entropy with energy for this particular form of perturbation.

We first consider the case when the momentum is a continuous random variable in Section~\ref{sec:proposedapproachcontinuous}. For the continuous case, by using the de Bruijn identity \cite{coverandthomas,stam} which provides a relationship
between rate of change of entropy with a parameter in the perturbation and the Fisher information, we show
that the thermodynamic temperature is the Fisher information associated with the probability density function of the momentum, i.e.,
and, hence, does not depend on form of the perturbation. The main result of this paper is that when the momenta are a vector valued random variable $\Pv$ with $3N$ components,
the thermodynamic temperature $\theta$ is given by
\begin{equation}
\frac{1}{\theta} = \frac{dS}{d {\cal E}} = \frac{2mk}{3N} \sum_{i=1}^{3N} \int_{-\infty}^{\infty} f_{\Pv}(\yv) \left(\frac{\partial}{\partial y_i} \ln f_{\Pv}(\yv) \right)^2 \ d \yv.
\end{equation}

Our definition of thermodynamic temperature coincides with the
conventional definition of temperature when the distribution is the Boltzmann distribution (steady-state distribution) but is more general since it is applicable to any distribution on the momentum.

In Section~\ref{sec:proposedapproachdiscrete}, we consider the case when the momentum is a discrete random variable and propose a particular form of perturbation that can be used to define the thermodynamic temperature. For this case also, we establish a relationship between the thermodynamic temperature and a statistical quantity associated with the distribution, which can be thought of as the discrete counterpart of Fisher information.

\section{Classical Definitions of Temperature}
\label{sec:classical}

Throughout the paper, we use the following notation: vector valued random variables are represented by capital letters with an arrow such as $\Pv$ and their realizations are denoted by lowercase letters with an arrow such as
$\yv$. The PDF of the random variable $\Pv$ evaluated at $\yv$ is denoted by $f_{\Pv}(\yv)$. Scalar random variables are denoted by capital letters and their realizations by lower case letters. Quantities associated with a random variable which are only functions of the PDF are denoted as functionals of the PDF instead of functions of the random variable. For example, the entropy of $\Pv$ is denoted by $S(f_{\Pv})$ instead of $S(\Pv)$. All logarithms considered are natural logarithms.

Consider a system with $N$ particles and let the
states be represented by the random variables $[\Xv,\Pv]$, where $\Pv \in R^{3N}$ denotes
the momenta and
$\Xv \in R^{3N}$ denotes the positions of the particles. Let $f_{\Xv \Pv}(\xv,\yv)$ be the joint probability density function of the
 position, $f_{\Xv}(\xv)$ and $f_{\Pv}(\yv)$ are the marginal PDF's of the position and momentum, i.e.,
 \[
 f_{\Xv}(\xv) = \int f_{\Xv \Pv}(\xv,\yv) \ d \yv, \ \ f_{\Pv}(\yv) = \int f_{\Xv \Pv}(\xv,\yv) \ d \xv.
 \]

In the above equation $d \yv = \displaystyle{\prod_{i=1}^{3N} dy_i}$ and $d \xv = \displaystyle{\prod_{i=1}^{3N} dx_i}$.

For an equilibrium distribution, we have the classical Boltzmann formula that
 \[
 f_{\Xv \Pv}(\xv,\yv) \propto e^{-\frac{1}{T} \left(\frac{||\yv||^2}{2m}+v(\xv) \right)},
 \]
where $\frac{||\yv||^2}{2m}+v(\xv)$ is the Hamiltonian of the system and $T$ is the temperature of the system.
Hence, the marginal
$f_{\Pv}(\yv)$ is a Gaussian distribution with zero mean and variance $2m {\cal E}(f_{\Pv})$, where
${\cal E}(f_{\Pv})$ is the kinetic energy of the system given by
 \begin{equation}
 {\cal E}(f_{\Pv}) = \frac{E[||\Pv||^2]}{2m} = \frac{1}{2m} \int_{-\infty}^{\infty} ||\yv||^2 \ f_{\Pv}(\yv) d \yv.
\label{eqn:energy}
 \end{equation}

The thermal entropy of the distribution $S$, is given by
\begin{equation}
S(f_{\Pv}) = - k \int f_{\Pv}(\yv) \ln (f_{\Pv}(\yv)) \ d \yv,
\label{eqn:shannonentropy}
\end{equation}
where $k$ is the Boltzmann constant.
At equilibrium, the temperature of the system is related to its kinetic
energy through the standard relationship

\begin{equation}
(3/2) N k T = {\cal E}(f_{\Pv}).
\label{eqn:kinetictemperature}
\end{equation}
For future reference, we will call $T$
as the kinetic temperature. Notice that $T$ is proportional to the variance of the momentum and, hence,
the kinetic energy of the system.

From a thermodynamic point of view, for a system in equilibrium,  an alternate definition of
temperature can be obtained through the fundamental equation of state \cite{callen}, from which we get
\begin{equation}
\frac{1}{\theta} = \frac{dS(f_{\Pv})}{d {\cal E}(f_{\Pv})},
\label{eqn:thermodynamictemperature}
\end{equation}
where $\theta$ is the thermodynamic temperature.

For a system in equilibrium, it well known that the kinetic temperature defined in (\ref{eqn:kinetictemperature})
is identical to the thermodynamic temperature defined in (\ref{eqn:thermodynamictemperature}), i.e. $\theta = T$.
Now, the question arises whether one can define these quantities for a system not in equilibrium, i.e., one
for which $f_{\Pv}(\yv)$ is not Gaussian. It is clear that the kinetic temperature can be defined exactly as in
(\ref{eqn:kinetictemperature}) for any distribution. For a general non-equilibrium distribution, if we adopt the
 Shannon entropy of a distribution as the equivalent of
Gibbs' entropy as is usually done, then there is no equation of state in general.
Hence, it is not possible to directly define $\frac{dS(f_{\Pv})}{d {\cal E}(f_{\Pv})}$ even though $S(f_{\Pv})$ and
${\cal E}(f_{\Pv})$ are well defined as in (\ref{eqn:shannonentropy}) and (\ref{eqn:energy}). Particularly, note that, away from
equilibrium $S(f_{\Pv})$ is not necessarily even a function of ${\cal E}(f_{\Pv})$.

\section{Perturbation Approach and Proposed Definition when Momentum is a Continuous Random Variable}
\label{sec:proposedapproachcontinuous}
Since both $S$ and ${\cal E}$ are functions of $f_{\Pv}$, we can treat the distribution $f_{\Pv}$ itself as a
parameter (i.e., we use the distribution $f_{\Pv}$ as descriptor of the macrostate of the system)
and intuitively define thermodynamic temperature as
\begin{equation}
\frac{dS(f_{\Pv})/df_{\Pv}}{d {\cal E}(f_{\Pv})/d f_{\Pv}}.
\end{equation}

In other words, given a probability distribution $f_{\Pv}$, we perturb it to $f_{\Pv'}$
{\footnote{Note that we use $f_{\Pv'}$ to denote the density function of the perturbed distribution instead of $f'_{\Pv}$.
It must be understood that the random variable $\Pv$ is perturbed to obtain a new random variable $\Pv'$ whose PDF is $f_{\Pv'}$.}}
and calculate the corresponding perturbations in $S$ and ${\cal E}$, namely $\Delta S$ and $\Delta {\cal E}$.
Then, we can define thermodynamic temperature as
\begin{equation}
\frac{1}{\theta} = \lim_{\Delta {\cal E} \rightarrow 0} \frac{\Delta S}{\Delta {\cal E}}.
\end{equation}

At a minimum, the temperature obtained by perturbing $f_{\Pv}$ should satisfy the following criteria
    \begin{enumerate}
    \item $\theta$ must be non-negative
    \item $\theta$ should be equal to the thermodynamic temperature for the Gaussian distribution
    \item $\theta$ should represent ``spread" of the kinetic energy, i.e., the more spread out the kinetic energy,
    the higher temperature
    \item $\theta$ should be a functional of the PDF $f_{\Pv}$
    \end{enumerate}

Not all perturbations of the probability distribution would give rise to sensible definitions
of temperature. In fact, it is quite possible to perturb the distribution in a way which will produce
unconventional results such as temperature being negative. The following example illustrates this

\underline{Example 1} For the sake of this example, consider a scalar random variable and
for any two $a$, $b$, such that $b > a > 0$, let $U_{[a,b]}(y)$ be the uniform distribution between $a$ and $b$, i.e.,
\begin{equation}
U_{[a,b]}(y) = \left\{
                 \begin{array}{ll}
                   \frac{1}{b-a}, & \hbox{$a \leq y \leq b$;} \\
                   0, & \hbox{otherwise.}
                 \end{array}
               \right.
\end{equation}
and let the probability distribution $f_P(y)$ be
\[
f_P(y) = \frac{1}{2} U_{[-b,-a]}(y) + \frac{1}{2} U_{[a,b]}(y).
\]

Consider a perturbation of $f_P(y)$ to $f_{P'}(y)$ given by
\[
f_{P'}(y) = \frac{1}{2} U_{[-\frac{b}{\gamma}-\Delta,-\frac{a}{\gamma}-\Delta]}(y) +
U_{[\frac{b}{\gamma}+\Delta,\frac{a}{\gamma}+\Delta]}(y).
\]

It can be seen that
\begin{eqnarray}
S(f_P) & = & \ln (b-a), \\
\label{eqn:gamma}
S(f_{P'}) & = & \ln \left(\frac{b-a}{\gamma}\right), \\
{\cal E}(f_P) & = & \frac{1}{2m}\frac{b^3-a^3}{3},\\
{\cal E}(f_{P'}) & = & \frac{1}{2m}\frac{(\frac{b}{\gamma}+\Delta)^3-(\frac{a}{\gamma}+\Delta)^3}{3}.
\end{eqnarray}

From (\ref{eqn:gamma}), it can be seen that when $\gamma = 1$, $\Delta S = S(f_{\Pv'}) - S(f_{\Pv}) = 0$.
However, for any $\Delta \geq 0$, $\Delta {\cal E} > 0$ and for this example, $\frac{\Delta S}{\Delta {\cal E}} = 0$.
By choosing appropriate values for $\gamma$ and $\Delta$, it is possible to get negative values of
$\frac{\Delta S}{\Delta {\cal E}}$. Thus demonstrating the fact that not perturbations are suitable
for a meaningful definition of temperature. We will now introduce a specific form of perturbation for which the temperature
defined in (\ref{eqn:thermodynamictemperature}) will satisfy conditions 1-4 mentioned in Section~\ref{sec:proposedapproachcontinuous}.

\subsection{Additive Perturbation}
From the macroscopic perspective, one can view the definition of
thermodynamic temperature in (\ref{eqn:thermodynamictemperature})
as the mathematical embodiment of the following thought experiment. We increase the total kinetic energy
of the particles by a small amount by ``heating" the system. Then, we measure the change in entropy of the
system. The ratio of the change in entropy to the change in energy is the inverse of the temperature. The
key point to observe here is that, this thought experiment depends upon the notion of heating the system
which guarantees both the entropy and energy increase (i.e., some sort of a diffusive process).
We now propose a statistical realization of this
notion.

Motivated by the kinetic theory interpretation of heating as due to the random collision of particles
with uncorrelated momenta, we consider an additive perturbation of the following form.
Let $P$ be the random variable which represents the momentum and consider a new
random variable $\Pv'$ given by
\begin{equation}
\Pv' = \Pv + \sqrt{\delta} \Qv,
\label{eqn:perturbation}
\end{equation}
where $\Qv$ is {\em any} random variable with zero mean and unit variance in each dimension and let
the components of $\Qv$ be independent of each other i.e., $E[\Qv \Qv^T] = I_{3N \times 3N}$.
Further, let $\Qv$ be {\em independent} of $\Pv$.
Let the PDF of $\Qv$ be $f_{\Qv}$ and that of
$\Pv'$ be $f_{\Pv'}$. Then, $f_{\Pv'}$ is
given by
\begin{equation}
f_{\Pv'}(\yv,\delta) = f_{\Pv}(\yv) \otimes f_{\sqrt{\delta} \Qv}(\yv),
\label{eqn:convolution}
\end{equation}
where $\otimes$ represents convolution. Note that $f_{\Pv'}(\yv,\delta)$ is explicitly a function
of $\delta$ also. Since $P$ and $Q$ are independent, the following two relations hold:

\begin{itemize}
\item[(i)] $S(f_{\Pv'})  \geq  S(f_{\Pv})$

This can be proved as follows.
Since conditioning cannot increase entropy \cite{coverandthomas},
\begin{equation}
S(f_{\Pv'}) \geq S(f_{\Pv'|\Qv}) = S(f_{\Pv+\Qv|\Qv}) = S(f_{\Pv})
\end{equation}

where $f_{\Pv'|\Qv}$ is the conditional distribution of $\Pv'$ given $\Qv$ and $S(f_{\Pv'|\Qv})$ is the entropy of $\Pv'$ given $\Qv$. The last equality follows
from the independence of $\Pv$ and $\Qv$. .

\item[(ii)] ${\cal E}(f_{\Pv'})  =  {\cal E}(f_{\Pv}) + \frac{3N}{2m} \delta$.

This follows from the fact that $\Qv$ is a random variable with zero mean which is independent of $\Pv$ and, hence,
$E[||\Pv'||^2] = E[||\Pv||^2] + E[(\sqrt{\delta} \Qv)^2] = E[||\Pv||^2] + \frac{3 N}{2 m} \delta $.
\end{itemize}

In other words, this method of perturbation is a way to increase the energy
by an known amount $\frac{3N\delta}{2m}$, which is also guaranteed to increase the entropy.
Furthermore, as will be seen later, $f_{\Pv'}(\yv,\delta)$ satisfies the diffusion
equation as $\delta \rightarrow 0$ in $3N$ dimensions. In other words, adding an independent random variable
to the momentum is equivalent to heating the body by a specified amount and as is
to be expected, it also increases the entropy. Hence, one would expect that the ratio
between the change in entropy associated with this perturbation to the change in
energy associated with this perturbation would be a measure of inverse temperature
of the distribution.

\paragraph{Proposed Definition of Temperature}
Hence, we formally define the inverse temperature of the system to be
\begin{equation}
 \frac{1}{\theta} = \lim_{\delta \rightarrow 0} \frac{S(f_{\Pv'})-S(f_{\Pv})}{{\cal E}(f_{\Pv'})- {\cal E}(f_{\Pv})} = \frac{2m}{3N} \lim_{\delta \rightarrow 0} \frac{S(f_{\Pv'})-S(f_{\Pv})}{\delta} = \frac{2m}{3N}\frac{\partial}{\partial \delta}
 S(f_{\Pv'}) |_{\delta = 0}
 \label{eqn:proposeddefinition}
\end{equation}

We will now show that the above quantity is independent of the actual perturbation
$f_{\Qv}$ in (\ref{eqn:perturbation}) and, hence, depends only on the distribution $f_{\Pv}$, which is intuitively pleasing. We also show that $1/\theta$ is the trace of the Fisher information matrix corresponding to the distribution of $f_{\Pv}$ with respect to the location family, scaled by $mk/3N$.

\subsection{Relationship between Temperature and Fisher Information}

Fisher information is a quantity that is commonly used in parametric estimation \cite{kullback}.
For a {\em scalar} random variable $P$ with a probability distribution $f_P(y)$, the Fisher information with respect to the location family, namely $J(f_P)$, is given by
\begin{equation}
J(f_P) = \int_{-\infty}^{\infty} f_P(y) \ \left[ \frac{\partial}{\partial y} \ln f_P(y) \right]^2 \ dy
= \int_{-\infty}^{\infty} f_P(y) \ \left[ \frac{\frac{\partial}{\partial y} f_P(y)}{f_P(y)} \right]^2 \ dy.
\label{eqn:fisherinformation1}
\end{equation}
Now, let us consider two distributions $f_P(y+t)$ and $f_P(y-t)$ which are shifted versions of $f_P(y)$,
shifted by $t$ to the left and right, respectively. Then, Fisher information can be expressed as
\begin{equation}
J(f_P) = \lim_{t \rightarrow 0} \frac{1}{t^2} \left( D(f_P(y+t) || f_{P}) + D(f_{{P}}(y-t) || f_P) \right)
\label{eqn:fisherrelativeentropy}
\end{equation}
where $D(f_P || g_{P})$ is the relative entropy (or Kullback-Leibler distance) between the distributions $f_P$ and $g_P$. This can be shown easily using the result in \cite{kullback}, where it is shown for a family of distributions $f_P(y;\omega)$ parametrized by $\omega$,
\[
\frac{1}{2}J(\omega_0) = \lim_{t \rightarrow 0} \frac{1}{t^2} D(f_P(y;\omega_0+t) || f_P(y;\omega_0))
\]
Now, considering two parametric families $f_P(y;\omega) = f_P(y-\omega)$ and $f_P(y;\omega) = f_P(y+\omega)$ and applying this result and taking the average we get the desired result. Note that for both these families $J(\omega_0) = J(f_P)$ as defined in (\ref{eqn:fisherinformation1}), which gives us the LHS of (\ref{eqn:fisherrelativeentropy}).

For a vector valued random variable $\Pv$ with $3N$ components, the $i,j$th entry of the Fisher information matrix is given by
\begin{equation}
\label{eqn:fisherinformationvector}
J_{i,j}(f_{\Pv}) = \int_{-\infty}^{\infty} f_{\Pv}(\yv) \left[ \frac{\partial}{\partial y_i} \ln f_{\Pv}(\yv) \frac{\partial}{\partial y_j} \ln f_{\Pv}(\yv)\right] d \yv, \ \ i,j = 1,\ldots,3N
\end{equation}


We now show that the temperature defined in (\ref{eqn:proposeddefinition}) is related to the trace of the Fisher information
matrix defined in (\ref{eqn:fisherinformationvector}), i.e., $\sum_i J_{i,i}(f_{\Pv})$.
The key result that we use to establish this connection is the
de Bruijn identity \cite{coverandthomas} which is given below for scalar random variables.

\begin{lemma}(de Bruijn Identity)
Let $P$ be a scalar random variable with a finite variance and let $f_P$ be the PDF of $P$. Let $Q$ be an independent random variable with unit variance and PDF $f_Q$, which is symmetric about 0, i.e., $f_Q(y) = f_Q(-y)$. Let $P'$ be the random variable given by
\begin{equation}
P' = P + \sqrt{\delta} Q.
\label{eqn:perturbation2}
\end{equation}

Then,
\underline{Arbitrary perturbation:} For {\em any} $f_Q$
\begin{equation}
\frac{\partial S(f_P')}{\partial \delta}|_{\delta = 0} = \frac{k}{2} J(f_P).
\label{eqn:debruijnarbitrary}
\end{equation}

\underline{Gaussian perturbation:} In the special case of $Q$ being a Gaussian random variable, the identity can be strengthened to
\begin{equation}
\frac{\partial S(f_{P'})}{\partial \delta} = \frac{k}{2} J(f_{P'}).
\label{eqn:debruijngaussian}
\end{equation}

\end{lemma}

The proof for the de Bruijn identity for the Gaussian perturbation can be found in \cite{coverandthomas}. However, the de Bruijn identity for an arbitrary perturbation is more relevant to us and a proof for this does not appear to be available in the literature (although the result appears to be known \cite{liunotes}). So, we prove the identity in (\ref{eqn:debruijnarbitrary}) here. Further, our proof also reveals some interesting characteristics of the perturbation which are discussed in Section~\ref{sec:heatequation}.

\underline{Proof:} Let $\phi_{P}(s) = \int_{-\infty}^\infty e^{sy} f_P(y) dy$ and $\phi_{Q}(s) = \int_{-\infty}^\infty e^{sy} f_{Q}(y) dy$ be the moment generating functions (MGF) of the random variables $P$ and $Q$ (i.e., $\phi_P(s)$ and $\phi_{Q}(s)$ are the Laplace transforms of $f_P(y)$ and $f_Q(y)$, respectively). The MGF of the random variable $\sqrt{\delta} Q$ is simply $\phi_{\sqrt{\delta} Q}(s) = \phi_Q(\sqrt{\delta} s)$.

For $P'$ given in (\ref{eqn:perturbation2}),
let $\phi_{P'}(s,\delta)$ be the MGF of $P'$ (note that we explicitly express the MGF as a function of $\delta$).
Since $P$ and $\sqrt{\delta} Q$ are independent, $\phi_{P'}(s,\delta)$ is given by
\begin{eqnarray}
\nonumber
\phi_{P'}(s,\delta) & = & \phi_{P}(s) \ \phi_{\sqrt{\delta} Q}(s) \\
& = & \phi_{P}(s) \left[ \int_{-\infty}^{\infty} e^{\sqrt{\delta} s y} f_Q(y) \ dy \right].
\label{eqn:MGF1}
\end{eqnarray}
Using a series expansion for $e^{\sqrt{\delta} s y}$, it can be seen that
\[
\int_{-\infty}^{\infty} e^{\sqrt{\delta} s y} f_Q(y) \ dy = \sum_{i=0}^\infty \frac{(\sqrt{\delta} s)^i}{i!} \ \mu_Q(i),
\]
where $\mu_Q(i)$ is the $i$th moment of $Q$. Since, we have assumed that $f_Q(-y) = f_Q(y)$, it can be readily seen that all the odd moments are zero and further, since we have
assumed $Q$ has unit variance, $\mu_Q(2)=1$. Therefore,
\[
\int_{-\infty}^{\infty} e^{\sqrt{\delta} s y} f_Q(y) \ dy  = 1 + \frac{\delta s^2}{2} + \sum \frac{\delta^k s^{2k}}{(2k)!} \mu_Q(2k).
\]
Substituting the above result into the right hand side of  (\ref{eqn:MGF1}), we get
\begin{equation}
\phi_{P'}(s,\delta)  = \phi_{P}(s) \left[1 + \frac{\delta s^2}{2} + \sum \frac{\delta^k s^{2k}}{(2k)!} \mu_Q(2k) \right].
\end{equation}
Taking the inverse Laplace transform on both sides, we get
\begin{equation}
f_{P'}(y,\delta) = f_{P}(y) + \frac{\delta}{2} \frac{\partial^2 f_P(y)}{\partial y^2} +
\sum \frac{\delta^k s^{2k}}{(2k)!} \mu_Q(2k)  \frac{\partial^{2k} f_P(y)}{\partial y^{2k}}.
\end{equation}

Differentiating the above equation with respect to $\delta$, we get

\begin{equation}
\frac{\partial f_{P'}(y,\delta)}{\partial \delta} = \frac{1}{2} \frac{\partial^2 f_P(y)}{\partial y^2} + o(\delta).
\end{equation}
Hence,
\begin{equation}
\frac{\partial f_{P'}(y,\delta)}{\partial \delta}|_{\delta=0} = \frac{1}{2} \frac{\partial^2 f_P(y)}{\partial y^2} = \frac{1}{2} \frac{\partial^2 f_{P'}(y)}{\partial y^2}|_{\delta = 0}.
\label{eqn:diffusion1}
\end{equation}

One can now follow the same proof as in \cite{coverandthomas} to prove the identity. Basically, we consider
the entropy of $P'$ namely
\begin{equation}
S(f_{P'}) = - k \int_{-\infty}^{\infty} f_{P'}(y,\delta) \ln f_{P'}(y,\delta) \ dy.
\end{equation}

Differentiating the above equation with respect to $\delta$, we get

\begin{eqnarray}
\nonumber
\frac{\partial S(f_{P'})}{\partial \delta} & = &- k \int_{-\infty}^{\infty} \left( \frac{\partial f_{P'}(y,\delta)}{\partial \delta} + \ln f_{P'}(y,\delta) \frac{\partial f_{P'}(y,\delta)}{\partial \delta} \right) dy \\
\nonumber
& = & - k \frac{\partial}{\partial \delta}\int_{-\infty}^{\infty} f_{P'}(y,\delta) dy - k \int_{-\infty}^{\infty}
\ln f_{P'}(y,\delta) \frac{\partial f_{P'}(y,\delta)}{\partial \delta} dy.
\end{eqnarray}

The first term can be seen to be zero since $\int_{-\infty}^{\infty} f_{P'}(y,\delta) dy=1$ and, hence,
\begin{eqnarray}
\frac{\partial S(f_{P'})}{\partial \delta} & = & - k \int_{-\infty}^{\infty}
\ln f_{P'}(y,\delta) \frac{\partial f_{P'}(y,\delta)}{\partial \delta} dy, \\
\Rightarrow \frac{\partial S(f_{P'})}{\partial \delta}|_{\delta = 0} & = & - k \int_{-\infty}^{\infty}
\ln f_{P'}(y,\delta) \frac{\partial f_{P'}(y,\delta)}{\partial \delta} dy |_{\delta = 0}.
\end{eqnarray}
Substituting (\ref{eqn:diffusion1}) in the above equation, we get
\begin{equation}
\frac{\partial S(f_{P'})}{\partial \delta}|_{\delta = 0} = - \frac{k}{2}  \int_{-\infty}^{\infty} \ln f_P(y) \frac{\partial^2 f_P(y)}{\partial y^2} \ dy.
\end{equation}
Now, integrating the RHS of the above equation by parts, we get
\begin{eqnarray}
\frac{\partial S(f_{P'})}{\partial \delta}|_{\delta = 0}  & = &
\frac{k}{2} \left[ - \ln f_P(y) \frac{\partial f_P(y)}{\partial y} \right]_{-\infty}^{\infty} +
k \int \frac{1}{f_P(y)} \left( \frac{\partial f_P(y)}{\partial y} \right)^2 \ dy \\
& = & \frac{k}{2} \left[ - \ln f_P(y) \frac{\partial f_P(y)}{\partial y} \right]_{-\infty}^{\infty} + \frac{k}{2} J(f_P).
\end{eqnarray}
The first term can be shown to be zero since it can be written as
$\frac{k}{2} \left[ \frac{\frac{\partial f_P(y)}{\partial y}}{\sqrt{f_P(y)}} 2 \sqrt{f_P(y)} \ln \sqrt{f_P(y)}\right]_{-\infty}^{\infty}$ and $\frac{\frac{\partial f_P(y)}{\partial y}}{\sqrt{f_P(y)}}$
is bounded since $\int_{-\infty}^{\infty} \left(\frac{\frac{\partial f_P(y)}{\partial y}}{\sqrt{f_P(y)}}\right)^2 dy = J(f_P)$, which is bounded. The second term is zero at both $y = \infty$ and $y = -\infty$ since $f_P(y) \rightarrow 0$ as $y \rightarrow \infty$ and $z \ln z \rightarrow 0$ as $z \rightarrow 0$. Hence, we get the desired result in the lemma. $\Box$

The relationship between thermodynamic temperature and Fisher information is given in the following
theorem.
\begin{thm}
\begin{equation}
\frac{1}{\theta} = \frac{2 m}{3N} \frac{\partial S(f_{\Pv'})}{\partial \delta}|_{\delta = 0} = \frac{m k}{3N} \sum_i J_{i,i}(f_{\Pv})
\end{equation}
\end{thm}

\underline{Proof:} Since the perturbation is independent in each dimension, we can apply Lemma 3.1 (de Bruijn identity)
to each component of the vector valued random variable, which gives the desired result. $\Box$

\underline{Example 2:} The difference between the kinetic temperature and the thermodynamic temperature is brought out in this example. Let the momentum be a scalar random variable $P$ whose probability density function given by
\[
f_P(y) = \frac{1}{\sqrt{2 \pi \sigma^2}} \frac{1}{2}
\{e^{-\frac{1}{2\sigma^2}(y-\mu)^2} + e^{-\frac{1}{2\sigma^2}(y+\mu)^2} \},
\]
i.e., $f_P(y)$ is the mixture of two Gaussian distributions with variance $\sigma^2$ and means
$\mu$ and $-\mu$. It is easy to see that the mean corresponding to $f_P(y)$ is zero and that the
energy is ${\cal E} = \mu^2 + \sigma^2$. Hence, the kinetic temperature for a given $\mu$ and $\sigma^2$ is
\[
T(\mu,\sigma^2) = \frac{1}{mk} (\mu^2+\sigma^2).
\]

The thermodynamic temperature however is related to the Fisher information associated with the distribution and is given by
\[
\theta(\mu,\sigma^2) = \frac{1}{mk} \frac{1}{J(f_P)},
\]
which can be numerically evaluated for a given $\mu$ and $\sigma^2$.

\begin{figure}
\begin{center}
  \includegraphics[width=4in,angle=0]{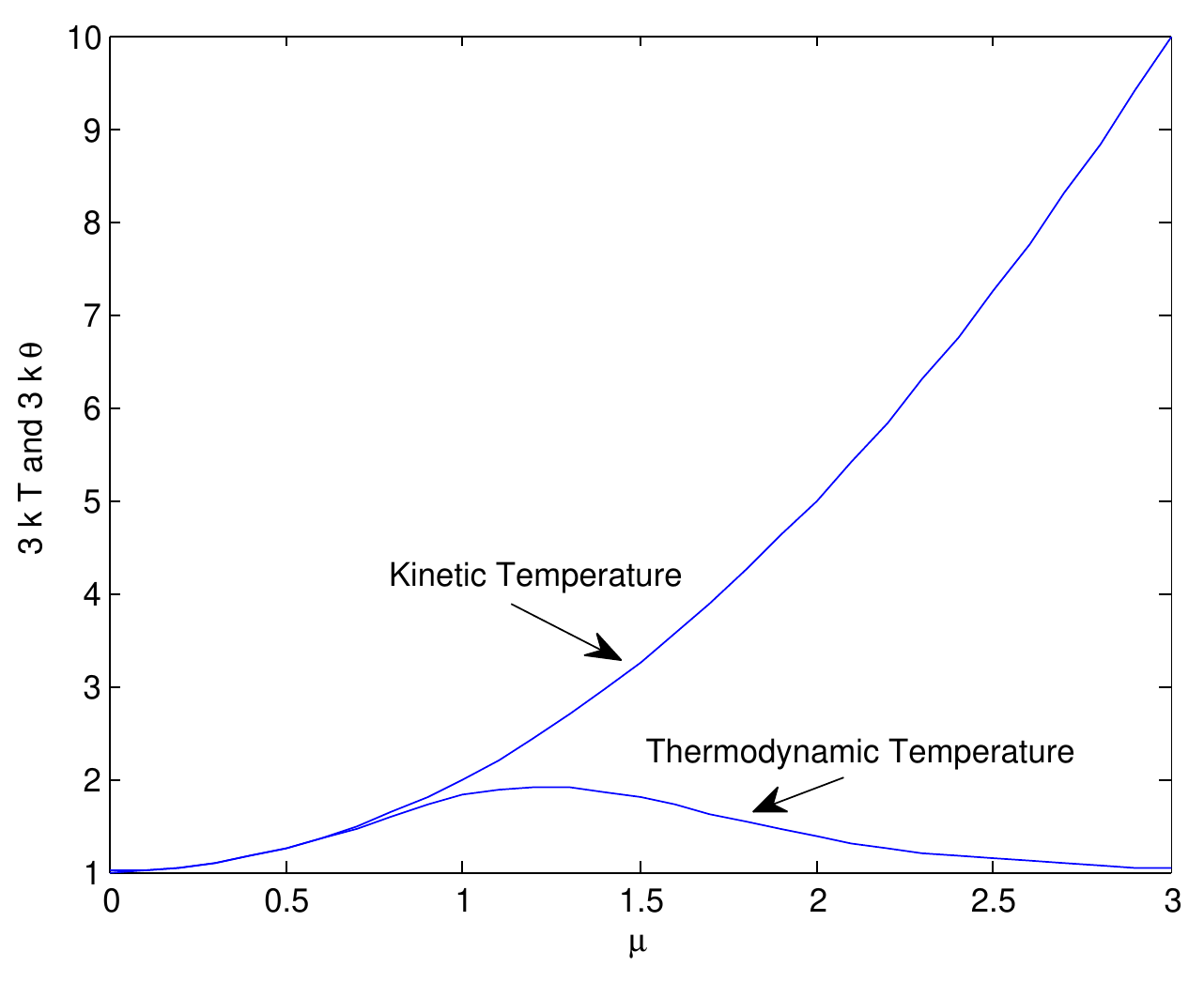}\\
  \caption{Kinetic and Thermodynamic temperature for a mixture Gaussian distribution}
\label{fig:example2}
\end{center}
\end{figure}

In Fig.~\ref{fig:example2}, we plot the kinetic temperature and thermodynamic temperature for $\sigma^2=1$ as
is $\mu$ varied. As $\mu$ increases, the distribution varies from a single Gaussian to a bimodal distributed composed of two Gaussians separated by a distance of $2\mu$. It can be seen that the kinetic temperature increases monotonically with $\mu$, whereas the thermodynamic temperature does not. A qualitative explanation of this phenomenon is provided by the observation that the thermodynamic temperature reflects the average ``local spread" of the distribution. Thus, when $\mu$ is close to zero, the thermodynamic temperature is close to the kinetic temperature. Whereas, when $\mu$ becomes large compared to $\sigma$, the distribution has two distinct peaks and locally has the same spread as that of a Gaussian. Thus, one would expect the thermodynamic temperature to return to its original value as $\mu$ increases, even though the average kinetic energy (and, hence, the kinetic temperature) increases monotonically with $\mu$.

\subsection{Remarks}
\label{sec:heatequation}

The above example illustrates the fact that unlike the equilibrium definition, the kinetic temperature and the thermodynamic temperature defined here show quite different behavior.  The difference is especially pronounced for
distributions such as for the bimodal distribution considered in the example. At this juncture, we note that a different notion of temperature has been introduced by Frieden \cite{frieden}, who defined a ``Fisher Temperature" as the {\em derivative} of the Fisher information associated with the given distribution with respect to any observable. Such a definition will give different values of Fisher temperature depending on the observable used.
On the other hand, the notion of thermodynamic temperature defined here is directly related to classical definitions, and uses the derivative of the classical entropy function with respect to energy. It is for this reason that we refer to the proposed definition as the ``thermodynamic" temperature.

Equation (\ref{eqn:diffusion1}) implies that regardless of the distribution of $Q$,
the newly added momentum diffuses through the system in the limit as $\delta \rightarrow 0$. It is this diffusion
of the added momentum that increases the entropy and the energy of the system as would be expected from any diffusion process.

\section{Perturbation and Proposed Definition when Momentum is a Discrete Random Variable}
\label{sec:proposedapproachdiscrete}
In this section, we consider the case when the momentum is a discrete random variable and for the sake of clarity,
we will consider only the scalar case. We assume that
the momentum $P$ can take on values $nh$, for any integer $n$ and a real constant $h$. Let $f_P$ denote
the probability mass function of $P$ and, for convenience let $f_P[n] = Pr(P = nh)$. There are two main reasons why the approach for the continuous case cannot be trivially extended to the discrete case. They are \begin{itemize}
\item In the discrete case, one cannot define an additive perturbation such as in (\ref{eqn:perturbation2}), since for arbitrary values of $\delta$, the perturbed random variable $P'$ will in general not be restricted to
the set $\{nh\}$
\item The definition of Fisher information in (\ref{eqn:fisherinformation1}) with respect to the location family requires probability density function $f_P$ to be smooth and, is hence, not directly applicable to discrete random variables.
\end{itemize}

We now propose a perturbation of the discrete random variable $P$ that results in the random variable $P'$ which also takes on values in $\{nh\}$. Let $Q$ be a discrete random variable also defined on $\{nh\}$ and let $f_Q[n] = Pr(Q = nh)$. Further, let $Q$ satisfy the following properties
\begin{itemize}
\item $f_Q$ is symmetric, i.e., $f_Q[-n] = f_Q[n]$
\item $f_Q[1] \neq 0$
\end{itemize}
For any $\delta > 0$, let us define $f_{Q,\delta}[n]$ as follows
\begin{equation}
f_{Q,\delta}[n] = \left\{
                    \begin{array}{ll}
                      f_Q[n] \delta^{|n|}, & n \neq 0; \\
                      1 - \sum_{n=1}^{\infty} (f_Q[n]+f_Q[-n]) \delta^n =
1 - \sum_{n=1}^{\infty} 2 f_Q[n] \delta^n , & n = 0.
                    \end{array}
                  \right.
\label{eqn:discreteperturbation1}
\end{equation}

\begin{figure}
\begin{center}
  \includegraphics[width=3.0in]{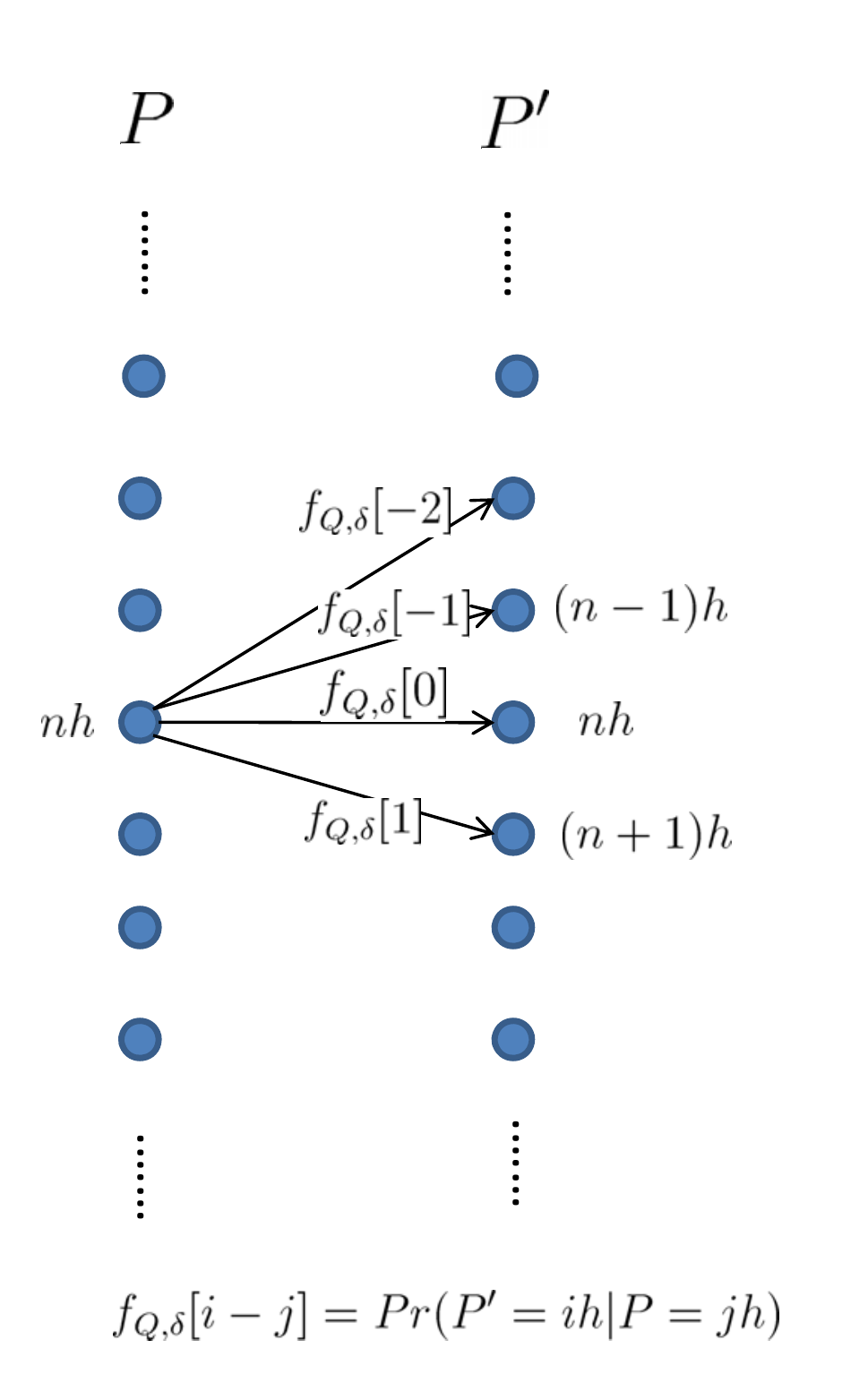}\\
  \caption{Equivalent communication channel between $P$ and $P'$}
\label{fig:communicationchannel}
\end{center}
\end{figure}

Now, consider a random variable $P'$ whose PMF $f_{P'}[n]$ is
\begin{equation}
f_{P'}[n,\delta] = f_P[n] \otimes f_{Q,\delta}[n],
\label{eqn:discreteperturbation2}
\end{equation}
where $\otimes$ refers to discrete convolution.
The random variable $P'$ can be thought as the output of a communication channel whose input
is $P$ and the transition probabilities in the communication channel are given by $f_{Q,\delta}[n]$
as shown in Fig.~\ref{fig:communicationchannel}. That is, $Pr(P' = ih | P = jh) = f_{Q,\delta}[i-j]$.

Let $H(f_P)  = -k \sum_{n=-\infty}^{\infty} f_P[n] \ln f_P[n]$ denote the entropy corresponding to the probability mass function $f_P$ (note that we use
$H$ instead of $S$, in accordance with standard notation in information theory for discrete random variables).
The perturbation in (\ref{eqn:discreteperturbation2}) can be shown to have the following properties
\begin{enumerate}
\item $H(f_{P'}) \geq H(f_P)$
\item ${\cal E}(f_{P'}) \geq {\cal E}(f_P)$
\end{enumerate}

The proof of Theorem 4.1 in the next section essentially proves these properties also. Since
this is developed in more detail in the next section, the proof is omitted here.

We now formally define the temperature as
\begin{equation}
\label{eqn:discretetemperature}
\frac{1}{\theta} = \lim_{\delta \rightarrow 0} \frac{H(f_{P'})-H(f_{P})}{{\cal E}(f_{P'}) - {\cal E}(f_P)}.
\end{equation}

We now show that this quantity can be expressed in terms of the relative entropies between $f_P[n]$, $f_P[n+1]$
and $f_P[n-1]$, where $f_P[n+1]$ and $f_P[n-1]$ refer to the distribution $f_P[n]$ shifted by one unit to the
left and right, respectively. We will first show the following lemma

\begin{lemma}
For any symmetric perturbation $f_Q$ and any probability density function $f_P$ such that $f_p[i] \neq 0, \forall i$,
\[
f_{P'}[n,\delta] = f_P[n] + \delta f_Q[1] (f_P[n-1]+f_P[n+1]-2) + o(\delta^2).
\]
\end{lemma}

\underline{Proof:} A Taylor's series expansion of $f_{P'}[n,\delta]$ with respect to $\delta$ around $\delta = 0$
gives
\[
f_{P'}[n,\delta] = f_P[n] + \delta \frac{\partial f_{P'}[n,\delta]}{\partial \delta}|_{\delta = 0} + o(\delta^2).
\]
To prove the lemma we only need to show that $\frac{\partial f_{P'}[n,\delta]}{\partial \delta}|_{\delta = 0}
= f_Q[1] (f_P[n-1]+f_P[n+1]-2)$. Let $\phi_P(s), \phi_{Q,\delta}(s)$ and $\phi_{P}(s,\delta)$ be the moment generating functions corresponding to the distributions $f_P[n]$, $f_{Q,\delta}[n]$ and $f_{P'}[n,\delta]$. Then,
\[
\phi_{P'}(s,\delta)  =  \phi_P(s) \phi_{Q,\delta}(s)
 =  \phi_P(s) \left( \sum_n e^{-snh} f_{Q,\delta}[n]\right)
\]
From (\ref{eqn:discreteperturbation1}), we get
\[
\phi_{Q,\delta}(s) = \sum_n e^{-snh} f_{Q,\delta}[n] = 1-2 f_Q[1] \delta - 2 \sum_{i=2}^\infty f_Q[i] \delta^i +
\left(e^{-sh}+e^{sh}\right) f_Q[1] \delta
\sum_{n=2}^\infty
\left(e^{-nsh}+e^{+nsh} \right) f_{Q}[n] \delta^n.
\]
Hence,
\begin{equation}
\nonumber
\phi_{P'}(s,\delta)  =  \phi_{P}(s) \left(1-2 f_Q[1] \delta - 2 \sum_{i=2}^\infty f_Q[i] \delta^i \right) +
\phi_P(s) \left(e^{-sh}+e^{sh} \right) f_Q[1] \delta
\phi_P(s) \sum_{n=2}^\infty
\left(e^{-nsh}+e^{+nsh} \right) f_{Q}[n] \delta^n.
\end{equation}

Grouping all the terms according to the exponents of $\delta$, we get

\begin{eqnarray}
\phi_{P'}(s,\delta)  & = & \phi_P(s) + \left(e^{-sh}+e^{sh}-2\right) f_Q[1] \delta \phi_P(s) + o(\delta^2), \\
\Rightarrow \frac{\partial \phi_{P'}[n,\delta]}{\partial \delta}|_{\delta = 0}
& = & \phi_P(s) \left(e^{-sh}+e^{sh}-2\right) f_Q[1].
\end{eqnarray}

Now taking the inverse Laplace transform, we get the desired result

\begin{equation}
\frac{\partial f_{P'}[n,\delta]}{\partial \delta}|_{\delta = 0}
= f_Q[1] (f_P[n-1]+f_P[n+1]-2).
\end{equation}

This results essentially means that, in the limit of $\delta \rightarrow 0$, it suffices to consider perturbations
for which only $f_Q[-1], f_Q[0]$, and $f_Q[1]$ are non-zero. Since $f_Q[-1] = f_Q[1]$ and $f_Q[-1]+f_Q[0]+f_Q[1]=1$, the
perturbation is of the form $f_Q[-1] = \gamma$, $f_Q[0] = 1-2 \gamma$ and $f_Q[1] = \gamma$.
We could have obtained this result without the use of the moment generating function by directly considering the convolution
of $f_P$ and $f_{Q,\delta}$ and then taking the limit of $\delta \rightarrow 0$. However, the use of the moment generating function
makes the derivation for the discrete case similar to that of the continuous case in Section~\ref{sec:proposedapproachcontinuous}.

We now show that our definition of temperature in (\ref{eqn:discretetemperature}) is closely related to the relative
entropies between the $f_P[n]$ and its shifted versions. This is made precise in the following theorem

\begin{thm}
Consider a perturbation $f_Q$ with $f_{Q}[-1]=f_Q[1]=\gamma$ and $f_Q[0]=1-2\gamma$. Then,
the inverse of thermodynamic temperature is
\begin{equation}
\frac{1}{\theta} = \lim_{\gamma \rightarrow 0} \frac{H(f_{P'})-H(f_P)}{{\cal E}(f_{P'}) - {\cal E}(f_P)} =
\frac{mk}{h^2} D(f_{P}[n+1] || f_{P}) + D(f_P[n-1] || f_P),
\end{equation}
\end{thm}
where $f_P[n-1]$ and $f_P[n+1]$ refer to the probability density function $f_P[n]$ shifted by one to the
right and left, respectively.

{\underline{Proof:}} For the perturbation under consideration  $f_{P'}$ is given by
\[
f_{P'}[n] = (1-2 \gamma) f_{P}[n] + \gamma (f_P[n-1] + f_P[n+1]).
\]

Let us first consider the difference in the energy
\begin{equation}
\nonumber
{\cal E}(f_{P'}) - {\cal E}(f_P) = \frac{1}{2m}\sum_n (nh)^2 (f_{P'}-f_P) = h^2 \sum_n n^2 \gamma (2 f_P[n] - f_P[n-1] - f_P[n+1]).
\end{equation}
Writing $\sum_n n^2 f_P[n-1]$ as $\sum_{n-1} ((n-1)^2 + 2n -1) f_P[n-1]$ and similarly, writing
$\sum_n n^2 f_P[n+1]$ as $\sum_{n+1} ((n+1)^2 + 2n -1) f_P[n+1]$ and simplifying, we get
\begin{equation}
\nonumber
{\cal E}(f_{P'}) - {\cal E}(f_P) = \frac{\gamma h^2}{m}.
\end{equation}

Now, let us consider the term $H(f_{P'}) - H(f_P)$:
\begin{eqnarray}
\nonumber
H(f_{P'}) - H(f_P) & = & -k \sum_n f_{P'}[n] \ln f_{P'}[n] + k \sum_n f_P[n] \ln f_P[n] \\
\nonumber
& = & - k \sum_n \left( (1-2 \gamma) f_{P}[n] + k \gamma (f_P[n-1] + f_P[n+1]) \right) \ln f_{P'}[n] + k \sum_n f_P[n] \ln f_P[n].
\end{eqnarray}
Since $H(f_P) = - k \sum_{n}f_{P}[n+1] \ln f_P[n+1] = - k \sum_{n}f_{P}[n-1] \ln f_P[n-1]$, we can add
$2H(f_P) + \sum_{n}f_{P}[n+1] \ln f_P[n+1] + \sum_{n}f_{P}[n-1] \ln f_P[n-1]$ to the right hand side without affecting
the result. Then, rearranging terms, we get
\begin{equation}
H(f_{P'}) - H(f_P) = k (1-2\gamma) D(f_P || f_{P'}) + k \gamma \left(D(f_P[n+1] || f_{P'}) +  k D(f_P[n-1] || f_{P'})\right).
\label{eqn:HPprimeminusHp}
\end{equation}

Writing a Taylor's series expansion for the first term $D(f_P || f_{P'})$ with respect to $\gamma$ about $\gamma = 0$, we
get
\begin{equation}
D(f_P || f_{P'}) = D(f_P || f_P) + \gamma \frac{\partial D(f_P || f_{P'})}{\partial \gamma}|_{\gamma = 0} + o(\gamma^2).
\end{equation}
Notice that $D(f_P || f_P) = 0$ and $\frac{\partial D(f_P || f_{P'})}{\partial \gamma}|_{\gamma = 0} = 0$ since
for a given $f_P$, $D(f_P || f_{P'})$ is continuous and convex in $f_{P'}$  with a minimum at $f_{P} = f_{P'}$ which
occurs that $\gamma = 0$ \cite{coverandthomas}. Hence, $D(f_P||f_{P'}) = o(\gamma^2)$ and, using this in (\ref{eqn:HPprimeminusHp}) and the fact that $\lim_{\gamma \rightarrow 0} f_{P'} = f_P$, we get
\begin{equation}
\lim_{\gamma \rightarrow 0} \frac{H(f_{P'})-H(f_P)}{{\cal E}(f_{P'}) - {\cal E}(f_P)} =  \frac{mk}{h^2} {D(f_P[n+1] || f_{P}) +  D(f_P[n-1] || f_{P})}.
\end{equation}

Note the similarity between this result and the relation in (\ref{eqn:fisherrelativeentropy}). In the limit as $h \rightarrow 0$, we recover the well known result in (\ref{eqn:fisherrelativeentropy}) for the continuous case.

\section{Conclusion}
In this paper, we have demonstrated that the notion of thermodynamical temperature can be extended to non-equilibrium distributions in a relatively straightforward way for both the discrete and continuous cases. In each situation, we introduce a perturbation which is ``diffusive".
A key point to note is that this definition of thermodynamical temperature retains all the features of the classical thermodynamic temperature without the need for any hypothesis of equilibrium. In other words, it is completely general. Although the ideas have been developed for the case of a system where the distribution is defined on the momenta alone, it is a trivial matter to extend it to joint distribution functions. Specifically, we can define a temperature field by using the condition distribution of the momentum given the position of the particle, i.e.,
$f_{\Pv|\Xv}$.


\end{document}